\documentclass[a4paper,12pt]{article}

\usepackage{amsmath}
\usepackage{amssymb}
\usepackage[dvips]{graphicx}
\usepackage[dvips]{psfrag}

\makeatletter
\@addtoreset{equation}{section}
\renewcommand{\theequation}{\thesection.\@arabic\c@equation}
\makeatother

\makeatletter
\renewcommand\appendix{\par
  \setcounter{section}{0}%
  \setcounter{subsection}{0}%
  \gdef\thesection{Appendix \@Alph\c@section }
  \renewcommand{\theequation}
  {\Alph{section}.\arabic{equation}}
}
\makeatother

\newcommand{\cO}{{\mathcal O}}

\newcommand{\bea}{\begin{eqnarray}}
\newcommand{\eea}{\end{eqnarray}}
\newcommand{\be}{\begin{equation}}
\newcommand{\ee}{\end{equation}}
\newcommand{\nn}{\nonumber}

\def\th{\theta}

\def\p{\partial}

  \def\cL{{\cal L}}
  \def\cO{{\cal O}}

\begin{document}

\titlepage

\vspace*{-15mm}
\baselineskip 10pt
\begin{flushright}
\end{flushright}
\baselineskip 24pt
\vglue 10mm

\begin{center}
{\Large\bf
Boundary Conditions for NHEK through\\ Effective Action Approach
}

\vspace{8mm}

\baselineskip 18pt

\renewcommand{\thefootnote}{\fnsymbol{footnote}}

Bin Chen$^{1,2}$\footnote[1]{\,bchen01@pku.edu.cn},
Bo Ning$^1$\footnote[2]{\,ningbo@pku.edu.cn}
and
Jia-ju Zhang$^1$\footnote[3]{\,jjzhang@pku.edu.cn}

\renewcommand{\thefootnote}{\arabic{footnote}}

\vspace{5mm}

{\it
$^{1}$Department of Physics, and State Key Laboratory of Nuclear Physics and Technology,\\
Peking University, Beijing 100871, P.R. China\\
\vspace{2mm}
$^{2}$Center for High Energy Physics, Peking University, Beijing 100871, P.R. China
}

\vspace{10mm}

\end{center}

\begin{abstract}
We study the asymptotic symmetry group(ASG) of the near horizon geometry of extreme Kerr black hole through the effective action approach developed in \cite{Porfyriadis:2010vg}. By requiring a finite boundary effective action, we derive a new set of asymptotic Killing vectors and boundary conditions, which are much more relaxed than the ones proposed in \cite{Matsuo:2009sj}, and still allow a copy of conformal group as its ASG. In the covariant formalism, the asymptotic charges are finite, with the corresponding central charge vanishing. By using the quasi-local charge and introducing a plausible cut-off, we find that the higher order terms of the asymptotic Killing vectors, which could not be determined through the effective action approach, contribute to the central charge as well. We also show that the boundary conditions suggested in \cite{Guica:2008mu} lead to a divergent first order boundary effective action.

\end{abstract}

\baselineskip 18pt

\newpage

\section{Introduction}\label{sec:Introduction}

The Kerr/CFT correspondence was first proposed  by studying the near horizon geometry of extreme Kerr black hole (NHEK) \cite{Guica:2008mu}, which has an $SL(2,\mathbb{R})\times U(1)$ isometry group. Following the spirit of \cite{Brown:1986nw}, the asymptotic symmetry group (ASG) of the NHEK geometry was studied. It was shown that the $U(1)$ isometry could be enhanced to a copy of Virasoro algebra, whose quantum version give rise to a central charge $c_L=12J$ with $J$ being the angular momentum of the black hole. From the Frolov-Thorne vacuum for extreme Kerr, it turns out that the dual CFT may have a nonvanishing left temperature $T_L=1/2\pi$. Then it was shown that the Bekenstein-Hawking entropy could be reproduced exactly by using the Cardy formula. This motivated the conjecture that the extreme Kerr black hole is dual to a two-dimensional CFT.

In the derivation of the ASG of NHEK, a set of boundary conditions was proposed in \cite{Guica:2008mu}.  These boundary conditions are unusual in the sense that some of the allowed deviations  are of the same order as the background. This is quite different from the  AdS$_3$ case studied in \cite{Brown:1986nw}, where all the deviations are subleading. The issue became more interesting when another set of consistent boundary conditions was proposed in \cite{Matsuo:2009sj}, from which it was shown that the $SL(2,\mathbb{R})$ isometry of NHEK could also be promoted to another Virasoro algebra.
This symmetry is related to the excitations of the right-moving sector \cite{Castro:2009jf}, which are suppressed in the extreme limit. The boundary conditions in \cite{Matsuo:2009sj} are similar to the usual ones, with all the deviations are subleading.
However the boundary conditions for  two copies of Virasoro algebra are not consistent with each other: the boundary conditions for the left mover excludes the right mover's, and vice versa. Although various efforts have been made \cite{Matsuo:2009yet,Matsuo:2010newlimit}, the boundary conditions simultaneously admitting two copies of Virasoro algebras are still unavailable.

Recently, a physical approach for \emph{deriving} the boundary conditions, rather than postulating them \emph{a priori}, was proposed by Porfyriadis and Wilczek \cite{Porfyriadis:2010vg}. By requiring finiteness of the boundary effective actions, they derived the asymptotic symmetries of new asymptotically AdS$_3$ spaces with further relaxed but still consistent boundary conditions. It is impressive that these asymptotic symmetries satisfy the Virasoro algebra with the same central charges  as the ones found in \cite{Brown:1986nw}. Despite its remarkable power in obtaining the ASG of AdS$_3$, it is not clear if the effective action approach is still efficacious in other cases.

In this paper, we would like to  apply this effective action approach to the study of the ASG and the consistent boundary conditions of the NHEK geometry. Although we fail to find the boundary conditions admitting two copies of Virasoro algebras, we do find  a set of 
new boundary conditions, which are more relaxed compared to the boundary conditions proposed in \cite{Matsuo:2009sj} but still allows one copy of conformal group as its ASG. In the covariant formalism, the asymptotic charges are finite, and the central charge turns out to be zero. As in \cite{Matsuo:2009sj}, we consider the quasi-local charge and obtain the corresponding central charge with appropriate regularization. We find that the central charge could not be determined since it depends on the higher order terms of the asymptotic Killing vectors, which could not be fixed in the effective action approach. Moreover, the anomalous transformations of the mass and the angular momentum depend on the higher order terms as well. This makes a consistent truncation on higher order terms impossible to account for a finite temperature effect. It turns out that to regain the expected right-moving central charge from the corresponding quantized Virasoro algebra, we have to revert to the boundary conditions presented in \cite{Matsuo:2009sj}. Furthermore our study shows that the boundary conditions suggested in \cite{Guica:2008mu} are in conflict with the finiteness of the boundary effective action. This indicates that the power of the effective action approach is restricted: it is not sufficient to fix the essential higher order terms of the ASG, in the meanwhile it excludes some of the interesting ASG's and consistent boundary conditions of a background.

\vspace*{2mm}

In the next section we review the effective action approach. In Section 3 we review briefly the NHEK geometry and its ASG discussed in the literature. We derive the asymptotic Killing vector $\xi$ and give the new boundary conditions for NHEK through effective action approach in Section 4. We end with some discussions in section 5. A discussion on asymptotic conformal Killing vectors is given in Appendix B.

\vspace*{2mm}

\section{Effective Action Approach}\label{sec:Action}

In \cite{Porfyriadis:2010vg}, Porfyriadis and Wilczek constructed  the effective action of general relativity (GR) for small excitations $g_{\mu\nu}\to g_{\mu\nu}+{\cal L}_\xi g_{\mu\nu}$ to the second order and derived the corresponding equation of motion. Starting from the Einstein-Hilbert action with a cosmological constant $\Lambda$,
\be\label{Einstein-Hilbert action}
S=\int_M d^n x\,\sqrt{-g}\,(R-2\Lambda)\,,
\ee
we put $g_{\mu\nu}\to g_{\mu\nu}+h_{\mu\nu}$ and expand to the second order in $h$:
\be\label{S0[h]+S1[h]+S2[h]}
S=S^{(0)}[h]+S^{(1)}[h]+S^{(2)}[h]+{\cal O}(h^3)\,.
\ee
The effective action for $\xi$ is obtained by putting $h_{\mu\nu}={\cal L}_\xi g_{\mu\nu}=\nabla_{\mu}\xi_{\nu}+\nabla_{\nu}\xi_{\mu}$. The first order action for $\xi$ turns out to be a boundary term:
\begin{eqnarray}\label{S1}
S^{(1)}[\xi]&=&\int_{\partial M} d^{n-1} x\,\sqrt{-\gamma}\,n^{\mu} \left(\Box\,\xi_{\mu}-\nabla_{\nu}\nabla_{\mu}\xi^{\nu}+(R-2\Lambda)\xi_{\mu}\right)\,,
\end{eqnarray}
which is a consequence of the diffeomorphism invariance of the Einstein-Hilbert action.

The second order action for $\xi$ takes the form
\begin{eqnarray}\label{S2}
S^{(2)}[\xi]&=&-\int_M d^n x\,\sqrt{-g}\,(G^{\mu\nu}+\Lambda g^{\mu\nu})(\xi_{\alpha;\,\mu\nu}-R_{\alpha\mu\nu\sigma}\xi^{\sigma})\,\xi^\alpha\nonumber\\
&&-\int_{\partial M} d^{n-1} x\,\sqrt{-\gamma}\,n^{\mu}\,\Big\{\xi_\sigma \nabla^\nu \nabla^\sigma\nabla_\nu \xi_\mu
-\xi_\mu \nabla^\nu\nabla^\sigma\nabla_\nu \xi_\sigma\\
&&\qquad\quad+(\nabla_\mu \xi_\nu+\nabla_\nu \xi_\mu)(\Box\, \xi^\nu-\nabla^\nu\nabla^\sigma \xi_\sigma)
+R_{\mu\nu\rho\sigma}\xi^\rho\nabla^\sigma \xi^\nu\nonumber\\
&&\qquad\quad+\frac{1}{2}\nabla_\mu\left[(\nabla^\nu \xi_\nu)^2-(\nabla^\nu \xi^\sigma)(\nabla_\nu \xi_\sigma)\right]
+2(\nabla^\nu \xi^\sigma)(\nabla_\nu\nabla_\sigma \xi_\mu-\nabla_\mu\nabla_\sigma \xi_\nu)\nonumber\\
&&\qquad\quad-\xi^\nu \nabla^\sigma\left[\xi_\mu (G_{\nu\sigma}+\Lambda g_{\nu\sigma})+\xi_\nu (G_{\mu\sigma}+\Lambda g_{\mu\sigma})-\xi_\sigma (G_{\mu\nu}+\Lambda g_{\mu\nu})\right] \Big\}\,.\nonumber
\end{eqnarray}
 If we assume that  the background $g_{\mu\nu}$ satisfies the Einstein field equation, $G^{\mu\nu}+\Lambda g^{\mu\nu}=0$, then the second order action also reduces to a boundary term, as a consequence of the gauge invariance of the second order action for $h$. While if we do not assume that the background solves the Einstein equation, the variational principle with $\delta \xi^\alpha$ leads to a beautiful equation of motion:
\be\label{EOM}
(G^{\mu\nu}+\Lambda g^{\mu\nu})(\xi_{\alpha;\mu\nu}-R_{\alpha\mu\nu\sigma}\xi^{\sigma})=0\,.
\ee
The equation (\ref{EOM}) is a contraction of two factors:
\begin{enumerate}
\item[i)] $G_{\mu\nu}+\Lambda g_{\mu\nu}=0\,$, which is satisfied by exact solutions to Einstein's GR;
\item[ii)] $\xi_{\alpha;\mu\nu}-R_{\alpha\mu\nu\sigma}\xi^{\sigma}=0\,$, which is satisfied by exact Killing vectors.
\end{enumerate}

The authors suggested that the Eq. (\ref{EOM}) should be satisfied in the asymptotic limit by approximate solutions and their corresponding asymptotic Killing vectors. Furthermore, it is expected to vanish fast enough so that the integrated action/unit time can be arbitrarily small provided one begin with $r$ sufficiently large. Most importantly, the boundary piece of $S^{(2)}[\xi]$ as well as the first order action $S^{(1)}[\xi]$ need to remain finite. It were such requirements that enable one to derive asymptotic symmetry vectors without imposing the boundary conditions \emph{a priori}.

\vspace*{2mm}

\section{The NHEK Geometry}\label{sec:NHEK}

Before applying the effective action approach to the NHEK case, let us review the NHEK geometry briefly. The Kerr metric in terms of the Boyer-Lindquist coordinates is of the form
\be
ds^2=-{\Delta \over \rho^2}\left(d\hat t-a \sin^2\theta d\hat\phi\right)^2+{\sin^2 \theta \over \rho^2}
\left((\hat r^2+a^2)d\hat \phi-a d\hat t\right)^2+
{\rho^2 \over\Delta}d\hat r^2+\rho^2 d\theta^2\,,
\ee
\be
\Delta\equiv\hat r^2-2M\hat r+a^2\:,\quad\quad\quad
\rho^2\equiv\hat r^2 +a^2\cos^2 \theta,\ee
where we take $G=\hbar=c=1$. It is parameterized by the mass $M$ and angular momentum $J=aM$. The horizons and the Hawking temperature are given by
\be
r_\pm=M\pm\sqrt{M^2-a^2}\,,\quad\quad\quad  T_H={r_+-M \over 4\pi Mr_+}\,.
\ee
We consider the near horizon geometry of the extreme Kerr with $J=M^2$. Defining new coordinates
\be
t=\frac{\lambda \hat{t}}{ 2M}\;, \;\;\;\;\;
x= \frac{\hat{r}-M }{\lambda M} \;, \;\;\;\;\;
\phi=\hat{ \phi}- {\hat{t} \over 2M}
\ee
and taking the limit $\lambda\to 0$, we find the NHEK geometry in Poincar\'{e}-type coordinates
\be
{ds^2 }= 2\, J\,\Gamma \left(-x^2dt^2+{dx^2\over x^2}+ d\theta^2 +\Omega^2  (d\phi + x dt)^2\right)\,
\ee
where
\be
\Gamma \equiv {1+\cos^2\theta\over 2}\;, \;\;\;\;\; \Omega \equiv {2 \sin \theta\over 1+\cos^2\theta}\,.
\ee
In global coordinates, we have
\be\label{nhek}
{d s^2 }= 2 \,J\,\Gamma \left( -(1+r^2){d\tau^2} + {dr^2\over 1+r^2} + d\theta^2 + {  \Omega^2}(d\phi + {rd\tau})^2\right).
\ee

In \cite{Guica:2008mu}, the following boundary conditions were chosen
\begin{equation}
 h_{\mu\nu} =
  \bordermatrix{
  &  &  &  &  \cr
  & \mathcal O(r^2)
  & \mathcal O(1/r^2)
  & \mathcal O(1/r)
  & \mathcal O(1)
  \cr
  &
  & \mathcal O(1/r^3)
  & \mathcal O(1/r^2)
  & \mathcal O(1/r)
  \cr
  &
  &
  & \mathcal O(1/r)
  & \mathcal O(1/r)
  \cr
  &
  &
  &
  & \mathcal O(1)
     }
\label{bdryAndy}
\end{equation}
in the basis ($\tau, r, \theta, \phi$). Here $h_{\mu\nu}$ is the deviation from the background (\ref{nhek}).
Note that the deviations $h_{\tau\tau}$ and $h_{\phi\phi}$ are of the same order as the leading terms in the background (\ref{nhek}).
The most general diffeomorphisms preserving the above boundary conditions, which requires
\be
\cL_\xi g_{\mu\nu}\sim h_{\mu\nu},
\ee  are
\be\label{asyAndy}
\xi=(-r\epsilon'_1(\phi)+\cO(1))\p_r+(C_1+\cO(1/r^3))\p_{\tau}+\cO(1/r)\p_\th+(\epsilon_1(\phi)+\cO(1/r^2))\p_\phi
\ee
where $\epsilon_1(\phi)$ is an arbitrary smooth function of $\phi$ and $C_1$ is an arbitrary constant. It does not contain the $SL(2,\mathbb{R})$ isometry subgroup of the background NHEK geometry, but still  contains a copy of the conformal group generated by
\be
\xi_1=\epsilon_1(\phi)\p_\phi-r\epsilon'_1(\phi)\p_r.
\ee
As $\phi$ is periodic, one may expand the function $\epsilon_1(\phi)$ and obtain a set of Virasoro generators. Furthermore, there is another translational symmetry generator $\p_\tau$, which commutes with $\xi_1$ and defines the energy $E$. As one takes the NHEK geometry as the ground state, one needs to impose the supplementary boundary condition $E=0$. It has been shown that the charge generating $\xi_1$ is finite around the NHEK geometry. Moreover, the Dirac bracket algebra of the charges gives rise to a central charge $c_L=12J$.

Another set of boundary conditions of NHEK was suggested in \cite{Matsuo:2009sj}
\begin{equation}
 h_{\mu\nu} =
  \bordermatrix{
  &  &  &  &  \cr
  & \mathcal O(1)
  & \mathcal O(1/r^3)
  & \mathcal O(1/r^3)
  & \mathcal O(1/r^2)
  \cr
  &
  & \mathcal O(1/r^4)
  & \mathcal O(1/r^4)
  & \mathcal O(1/r^3)
  \cr
  &
  &
  & \mathcal O(1/r^3)
  & \mathcal O(1/r^3)
  \cr
  &
  &
  &
  & \mathcal O(1/r^2)
  \cr
  }
\label{bdryMatsuo}
\end{equation}
in the basis ($\tau, r, \theta, \phi$). Obviously all the deviations are subleading.
The most general form of the asymptotic diffeomorphism preserving these boundary conditions is
\bea
\label{asyMatsuo}
 \xi_2&=&\left(-r\epsilon'(\tau)+\frac{\epsilon'''(\tau)}{2r}+\cO(1/r^2)\right)\p_r+\left(\epsilon(\tau)
 +\frac{\epsilon''(\tau)}{2r^2}+\cO(1/r^3)\right)\p_{\tau} \nn\\
 & &+\cO(1/r^3)\p_\th+\left(C_2-\frac{\epsilon''(\tau)}{r}+\cO(1/r^3)\right)\p_\phi
\eea
where $\epsilon(\tau)$ is an arbitrary function of $\tau$ and $C_2$ is a constant. It contains all of the isometries of the NHEK geometry. In particular the $SL(2,R)$ symmetry could be enhanced to the Virasoro algebra. However, the algebra of the corresponding asymptotic charge\cite{Barnich:2001jy, Barnich:2007bf}  do not have a central extension. Nevertheless, using the quasi-local charge instead gives a nonvanishing central charge, which depends on a cutoff. This nonvanishing central charge suggest that there is a right-moving sector with physical degrees of
freedom.

\section{Asymptotic Killing Vectors for NHEK}\label{sec:AKV}

In this section we derive the asymptotic Killing vectors $\xi$ for the NHEK geometry through the effective action approach. We assume a power series expansion of the components of  $\xi$ as
\be\label{expansion}
\xi^\mu=\sum_{n \in Z} \xi^\mu_n (\tau,\theta,\phi)\, r^n\,
\ee
and assume that each series truncates for some large $N$ onwards.

According to Porfyriadis and Wilczek \cite{Porfyriadis:2010vg}, the asymptotic symmetry algebra generating vectors $\xi$ are obtained by requiring ``small" asymptotic transformation, where ``smallness" is defined by requiring subleading Lie derivatives of the background metric, finite first order effective action $S^{(1)}[\xi]$, and finite second order effective action $S^{(2)}[\xi]$. For the NHEK case, since the boundary conditions with leading order perturbations had been exhibited in \cite{Guica:2008mu}, it is reasonable for us to allow the leading order Lie derivatives of the NHEK so as to accommodate all possibilities. We require that the ${\cal L}_\xi g_{\mu\nu}$ are of the leading order for the non-vanishing NHEK components $g_{\mu\nu}$ in (\ref{nhek}) and finite for others, that is
\be
{\cal L}_{\xi} g_{\mu\nu} \sim h^0_{\mu\nu},
\ee
where
\begin{equation}
 h^0_{\mu\nu} =
  \bordermatrix{
  &  &  &  &  \cr
  & \mathcal O(r^2)
  & \mathcal O(1)
  & \mathcal O(1)
  & \mathcal O(r)
  \cr
  &
  & \mathcal O(1/r^2)
  & \mathcal O(1)
  & \mathcal O(1)
  \cr
  &
  &
  & \mathcal O(1)
  & \mathcal O(1)
  \cr
  &
  &
  &
  & \mathcal O(1)
  \cr
  }
\label{bdry0}
\end{equation}
in the basis ($\tau, r, \theta, \phi$). The most general $\xi'$s satisfying the above conditions are given by:
\begin{eqnarray}
\xi^\tau&=&\epsilon(\tau)+\xi^\tau_{-1}(\tau,\phi)\frac{1}{r}+{\cal O}(\frac{1}{r^2})\,,\nn\\
\xi^r&=&{\cal O}(r)\,,\label{xi1}\\
\xi^\theta&=&{\cal O}(1)\,,\nn\\
\xi^\phi&=&\xi^\phi_{0}(\tau,\phi)+{\cal O}(\frac{1}{r})\,,\nn
\end{eqnarray}
where $\xi^\tau_{-1}$ and $\xi^\phi_{0}$ are arbitrary functions of $(\tau, \phi)$, and  $\epsilon(\tau)=\xi^\tau_0$ is an arbitrary function of $\tau$. It accommodates both the diffeomorphisms (\ref{asyAndy}) and (\ref{asyMatsuo}).

Perturbing the NHEK metric (\ref{nhek}) using the $\xi'$s in (\ref{xi1}) gives the following boundary conditions
\begin{equation}
 h^1_{\mu\nu} =
  \bordermatrix{
  &  &  &  &  \cr
  & \mathcal O(r^2)
  & \mathcal O(1)
  & \mathcal O(1)
  & \mathcal O(r)
  \cr
  &
  & \mathcal O(1/r^2)
  & \mathcal O(1/r)
  & \mathcal O(1/r)
  \cr
  &
  &
  & \mathcal O(1)
  & \mathcal O(1)
  \cr
  &
  &
  &
  & \mathcal O(1)
  \cr
  }.
\label{bdry1}
\end{equation}
These boundary conditions are so relaxed that they accommodate both (\ref{bdryAndy}) and (\ref{bdryMatsuo}).

In order that the integrand of the first order effective action $S^{(1)}[\xi]$ (\ref{S1}) is finite everywhere on the boundary $r=\infty$, we find that the following relations among the leading order terms as well as the next-to-leading order terms in (\ref{xi1}) must be satisfied:
\bea
&&\xi^r_1~~~~~\,\,=\,-\xi^\tau_{0,\,\tau}-\xi^\phi_{0,\,\phi}\,,\label{relation1}\\
&&\xi^\tau_{-1,\,\phi}~~=-2\,\xi^\phi_{0,\,\phi}\,,\label{relation2}\\
&&\xi^\phi_{0,~\phi\phi\phi}\,=\,0\,.\label{relation3}
\eea
Note that the equations (\ref{relation2}) and (\ref{relation3}) exclude explicitly the diffeomorphisms (\ref{asyAndy}) given in \cite{Guica:2008mu}. 
In other words, the asymptotic transformations (\ref{asyAndy}) would lead to divergent first order boundary effective action. 

Since the NHEK metric (\ref{nhek}) is also an exact solution of the Einstein field equation with vanishing cosmology constant, the second order effective action $S^{(2)}[\xi]$ (\ref{S2}) is again a boundary term. We require that the integrand of  $S^{(2)}[\xi]$ to be finite as before. This imposes
\bea
&&\xi^\tau_{-1}\,=0\,,\\
&&\xi^\theta_{0,\,\phi}=0\,
\eea
at leading order and a complicated equation at next-to-leading order, which is shown in Appendix A. In order to find a particular solution, we simply assume that $\xi^\theta_0=0$, and $\xi^\phi_{-1}\,,~\xi^\tau_{-2}\,,~\xi^r_{0,\,\phi}\,,~\xi^\theta_{-1,\,\phi}$ are all proportional to $\epsilon''(\tau)$, then we get the following solution
\be
\xi^\phi_{-1}=-\,\epsilon''(\tau)\,,\quad\quad \xi^\tau_{-2}=\frac{1}{2}\,\epsilon''(\tau)\,, \quad\quad
\xi^r_{0,\,\phi}=0\,,\quad\quad \xi^\theta_{-1,\,\phi}=0\,.\label{partsl}
\ee
We therefore arrive at
\begin{eqnarray}
\xi^\tau&=&\epsilon(\tau)+\epsilon''(\tau)\frac{1}{2r^2}+{\cal O}(\frac{1}{r^3})\,,\nn\\
\xi^r&=&-r\epsilon'(\tau)+\xi^r_0(\tau,\theta)+{\cal O}(\frac{1}{r})\,,\label{xi2}\\
\xi^\theta&=&\xi^\theta_{-1}(\tau,\theta)\frac{1}{r}+{\cal O}(\frac{1}{r^2})\,,\nn\\
\xi^\phi&=&\xi^\phi_0(\tau)-\epsilon''(\tau)\frac{1}{r}+{\cal O}(\frac{1}{r^2})\,.\nn
\end{eqnarray}
The $\xi$'s in (\ref{xi2}) are our final asymptotic Killing vectors of NHEK. The $\xi^\tau$ and the leading term of  $\xi^r$ as well as the subleading term of $\xi^\phi$ are exactly the same as those in Eq. (\ref{asyMatsuo}). However,
the subleading term of $\xi^r$ and the leading term of $\xi^\phi$ are much more relaxed, so as all of the terms of $\xi^\theta$. As shown in \cite{Matsuo:2009sj}, expanding into modes $\epsilon(\tau)=\tau^{1+n}$,
one finds that to the leading order in $1/r$, the generators $\xi_n$ form the Virasoro algebra
\begin{equation}
 [\xi_n, \xi_m]_{\text{L.B.}} = {\cal L}_{\xi_n}\xi_m = (m-n)\,\xi_{m+n} \,,\label{virasoro}
\end{equation}
here for the closing of the Virasoro algebra we define
\bea
\xi^r_{0\,(m+n)} &=& \frac{1}{m - n}\left((1 + n)\, \tau^n \,\xi^r_{0\,(m)} - (1 + m)\, \tau^m\, \xi^r_{0\,(n)}  + \tau^{n+1}\, \xi^r_{0\,(m),\,\tau}
 - \tau^{m+1}\, \xi^r_{0\,(n),\,\tau}\right) \,,\nn\\
\xi^\theta_{-1\,(m+n)} &=& \frac{1}{m - n}\left((1 + n)\, \tau^n \,\xi^\theta_{-1\,(m)} - (1 + m)\, \tau^m\, \xi^\theta_{-1\,(n)}
+ \tau^{n+1}\, \xi^\theta_{-1\,(m),\,\tau} - \tau^{m+1}\, \xi^\theta_{-1\,(n),\,\tau}\right) \,,\nn\\
\xi^\phi_{0\,(m+n)} &=& \frac{1}{m - n}\left(\tau^{n+1}\, \xi^\phi_{0\,(m),\,\tau} -  \tau^{m+1}\, \xi^\phi_{0\,(n),\,\tau}\right)\nn\,.
\eea

Note that for (\ref{virasoro}) we do not need to require $\xi^\phi_0(\tau)=0$. This is more relaxed compared to the case of  \cite{Matsuo:2009sj}, where the leading term of $\xi^\phi$, i.e. the constant $C_2$ of (\ref{asyMatsuo}) need to be zero in order that the Virasoro algebra is closed.

A discussion on asymptotic conformal Killing vectors of NHEK which leave the first order effective action finite is given in Appendix B.

\vspace*{2mm}


Having established the asymptotic Killing vectors (\ref{xi2}) through the effective action approach, new boundary conditions could be obtained by perturbing the exact NHEK metric (\ref{nhek}) using these $\xi$'s\,:
\begin{equation}
 h_{\mu\nu} =
  \bordermatrix{
  &  &  &  &  \cr
  & \mathcal O(r)
  & \mathcal O(1/r^2)
  & \mathcal O(1/r)
  & \mathcal O(1)
  \cr
  &
  & \mathcal O(1/r^3)
  & \mathcal O(1/r^2)
  & \mathcal O(1/r^3)
  \cr
  &
  &
  & \mathcal O(1/r)
  & \mathcal O(1/r^2)
  \cr
  &
  &
  &
  & \mathcal O(1/r)
  \cr
  }.
\label{bdry2}
\end{equation}
 These boundary conditions are more relaxed than the ones (\ref{bdryMatsuo}) but still subleading compared to the background (\ref{nhek}).

We  check that the Lie derivatives of the perturbations still satisfy the boundary conditions, i.e. ${\cal L}_{\xi} h_{\mu\nu} \sim h_{\mu\nu}$.
Furthermore, adopting the covariant formalism developed by Barnich, Brandt and Comp\`ere \cite{Barnich:2001jy, Barnich:2007bf}, we find that the asymptotic charges $Q[\xi]$ corresponding to (\ref{xi2}) and (\ref{bdry2}) are finite, with the trivial symmetry transformations
\bea
 \xi_{tr}&=&\cO(\frac{1}{r^3})\p_{\tau} +\left(\xi^r_0(\tau,\theta)+\cO(\frac{1}{r})\right)\p_r\nn\\
&& +\left(\xi^\theta_{-1}(\tau,\theta)\frac{1}{r}+\cO(\frac{1}{r^2})\right)\p_\theta
 +\left(\xi^\phi_0(\tau)+\cO(\frac{1}{r^2})\right)\p_\phi  \label{xitr}
\eea
give rise to vanishing charges. The central charge of the Virasoro algebra at the Dirac bracket level turns out to be zero.

However, as in \cite{Matsuo:2009sj} we could also employ the quasi-local charge\cite{Brown:1992br} defined by using the surface energy momentum tensor to study the central charge. Unfortunately, it turns out that the terms in (\ref{xitr}) contribute to the central charge. We find that the following leading terms of the asymptotic Killing vectors
\be
 \xi_{\,l}=\left(\epsilon(\tau)+\epsilon''(\tau)\frac{1}{2r^2}\right)\p_{\tau} - r \,\epsilon'(\tau) \p_r
-\epsilon''(\tau)\frac{1}{r}\, \p_\phi  \label{xil}
\ee
give rise to Virasoro algebra with central charge $c=6\, a^2/(G \Lambda)$, half of the value  obtained in \cite{Matsuo:2009sj}. Moreover, when calculating the anomalous transformations of the mass and the angular momentum, we get
\bea
\delta M &=& -\frac{a^2}{2 G \Lambda}\epsilon'''(\tau)\,,\nn\\
\delta J &=& -\frac{3 a^2}{2 G \Lambda^2}\epsilon'''(\tau)\,.
\eea
where $\Lambda$ is a large but finite radius where the boundary locates.
Comparing with the results  in \cite{Matsuo:2009sj}, we find that our $\delta M$ is again half of the value obtained there, while $\delta J$ is one and a half of the value. Following the analysis of finite temperature effects in \cite{Matsuo:2009sj}, we find that it is impossible to match the entropy and the mass as well as the angular momentum between the Kerr black hole and the boundary CFT simultaneously. However, if we add a term
\be
\xi_{ad}=\frac{\epsilon'''(\tau)}{2\,r}\,\p_r
\ee
to (\ref{xil}), we recover the same central charge and anomalous transformations as those in \cite{Matsuo:2009sj}. This shows that the higher order terms which could not be determined by the current effective action approach play essential roles. In fact all of the undetermined terms in (\ref{xitr}) contribute to the anomalous transformations as well as the central charge (see Appendix C). To regain the results obtained in \cite{Matsuo:2009sj}, the simplest way is to set the coefficient $\xi^r_{-1} = \epsilon'''(\tau)/2 $ and all of the other terms in (\ref{xitr}) to be zero, reverting to the ASG as well as the boundary conditions proposed in \cite{Matsuo:2009sj}.

\vspace*{2mm}

\section{Discussion}\label{discussion}

In this paper we derived new boundary conditions for the NHEK geometry through the effective action approach. We found that both the boundary conditions and the corresponding asymptotic Killing vector are relaxed compared to the ones in \cite{Matsuo:2009sj}. These boundary conditions are consistent and of subleading order asymptotically, leading to finite charges. However, although the corresponding ASG contains a copy of conformal group, their generators give different right-moving central charge when being quantized. To recover the result obtained in \cite{Matsuo:2009sj}, we have to restrict our asymptotic Killing vectors and eventually go back to the ASG as well as the boundary conditions in \cite{Matsuo:2009sj}, which could not be derived through the effective action approach. This shows that the power of the effective action approach is restricted, at least for the NHEK case.

The reason for this issue might be that the effective action (\ref{S0[h]+S1[h]+S2[h]}) developed by Porfyriadis and Wilczek was constructed only to the second order. In fact in their paper \cite{Porfyriadis:2010vg}, although more relaxed asymptotic Killing vectors for the AdS$_3$ were obtained, giving rise to the correct central charge, the higher order terms of the vectors (21) did contribute to the central charge. The point is that the contributions from the higher order terms of Brown-Henneaux's asymptotic symmetries (3) canceled each other totally, leading to the same central charge. It is possible that with higher order expansions of the action, we could be able to determine the essential higher order terms of the asymptotic Killing vectors (\ref{xi2}). 

Moreover, it is quite unexpected that the effective action approach is not consistent with the diffeomorphisms in \cite{Guica:2008mu}. In fact one can check straightforwardly that the diffeomorphisms subject to the boundary conditions in \cite{Guica:2008mu} lead to a divergent first order boundary effective action. It may be general that the effective action approach does not allow for the boundary conditions with leading order perturbations, therefore excludes a variety of interesting possibilities of consistent boundary conditions.

On the other hand, our search for the ASG and consistent boundary conditions is not exhaustive. Note that our asymptotic Killing vector is only a particular solution of the equations from the finiteness of the second order effective action. It is possible that there exist more general solutions which lead to different consistent boundary conditions.

\vspace*{10mm}

\noindent
 {\large{\bf Acknowledgments}}

 The work was in part supported by NSFC Grant No. 10775002, 10975005. BC would like to thank
 the organizer and participants of the advance workshop ``Dark Energy and Fundamental Theory" supported by
 the Special Fund for Theoretical Physics from the National Natural Science Foundations of China with grant no: 10947203 for stimulating discussions and comments. BN would like to thank the hospitality of the National Center for Theoretical Sciences, Taipei, Taiwan, during the final stage of the work.

\vspace*{2mm}

\appendix

\section{Equation from Second Order Action}

Requiring the integrand of $S^{(2)}[\xi]$ (\ref{S2}) to be finite give rise to the following complicated equation at next-to-leading order:
\bea
0&=& 32 (3 + \cos{2\theta}) \epsilon'(\tau)
(16 (-123 \cos{\theta} + 29 \cos{3\theta} - 37 \cos{5\theta} + 3 \cos{7\theta}) \xi_0^\theta\nn\\
&&   + 2 (5 \sin{\theta} + \sin{3\theta}) (32 \cos^2{\theta} (9 - 5 \cos{2\theta}) \xi^\theta_{0,\,\theta} - 2 (5 \sin{\theta} + \sin{3\theta}) ((13 \cos{\theta} + 3 \cos{3\theta}) \xi^\theta_{0,\,\theta\theta}\nn\\
&&+ (5 \sin{\theta} + \sin{3\theta})\xi^\theta_{0,\,\theta\theta\theta})))  - 2 \sin{\theta}\, \epsilon(\tau) (8192 (-20 \cos{2\theta} +
3 (-5 + \cos{4\theta})) \sin^2{\theta}\, \xi^\phi_{-1}  \nn\\
&& -128 (2579 + 3616 \cos{2\theta} - 84 \cos{4\theta} + 32 \cos{6\theta} +
\cos{8\theta}) \sin^2{\theta}\, \xi^\tau_{-2} \nn\\
&&  + 2 (3 + \cos{2\theta}) (64 (86 + 143 \cos{2\theta} +
            26 \cos{4\theta} + \cos{6\theta}) \sin^2{\theta}\, \epsilon''(\tau)\nn\\
&&+ 16 (  (-1250 \sin{2\theta} + 842 \sin{4\theta} +
               22 \sin{6\theta} + 3 \sin{8\theta}) \xi^\tau_{-2,\,\theta}\nn\\
&&  + 16 \cos{\theta}\, \sin^3{\theta} \,(2 (51 + 12 \cos{2\theta} + \cos{4\theta}) \xi^\theta_{0,\,\tau}
- (83 + 12 \cos{2\theta} + \cos{4\theta}) \xi^\theta_{-1,\,\phi} \nn\\
&& + 8 (-21 + \cos{2\theta}) \xi^\phi_{-1,\,\theta}))+ 2 (3 + \cos{2\theta}) (64 (3 + 28 \cos{2\theta} + \cos{4\theta}) \sin^2{\theta}\, \xi^r_{0,\,\phi}\nn\\
&&- 32 (3 + 28 \cos{2\theta} + \cos{4\theta}) \sin^2{\theta} \,\xi^\phi_{-1,\,\phi\phi}
 - (384  - 512 \cos{2\theta}  + 128 \cos{4\theta}) \xi^\phi_{-1,\,\theta\theta}\nn\\
&& + 64 (3 + 28 \cos{2\theta} + \cos{4\theta}) \sin^2{\theta} \,\xi^\tau_{-2,\,\theta\theta}\nn\\
& &+ (803  + 392 \cos{2\theta}  + 796 \cos{4\theta}  + 56 \cos{6\theta}
 + \cos{8\theta}) \,\xi^\tau_{-2,\,\phi\phi}\,)))
\eea
Assuming that $\xi^\theta_0=0$, and $\xi^\phi_{-1}\,,~\xi^\tau_{-2}\,,~\xi^r_{0,\,\phi}\,,~\xi^\theta_{-1,\,\phi}$ are all proportional to $\epsilon''(\tau)$ with the coefficients $A_1$, $A_2$, $A_3$, $A_4$, the above equation is greatly simplified to the following form:
\bea
0&=&659 - 1920 A_1 - 5158 A_2 + 451 A_3  +\,(1056 - 2560 A_1 - 7232 A_2 + 1176 A_3) \cos{2\theta} \nn\\
&& +\, (300 + 384 A_1 + 168 A_2 + 380 A_3) \cos{4\theta}  +\, (32 - 64 A_2 + 40 A_3) \cos{6\theta} \nn\\
&& +\, (1 - 2 A_2 + A_3) \cos{8\theta} -\, A_4 \,(1002 \sin{2\theta} + 238  \sin{4\theta} + 18  \sin{6\theta} +  \sin{8\theta}). \label{simpl}
\eea
The solution to Eq. (\ref{simpl}) is of the form (\ref{partsl}).

\section{Asymptotic Conformal Killing Vectors}
\label{sec:Conformal}

In this appendix we work out the asymptotic conformal Killing vectors of NHEK which leave the first order effective action finite. Assume that the vector $\xi$ satisfies the conformal Killing equation for NHEK in the asymptotic limit $r\to\infty$,
\be\label{ckv}
\nabla_\mu\xi_\nu+\nabla_\nu\xi_\mu=\frac{1}{2}g_{\mu\nu}\nabla_\sigma\xi^\sigma\,.
\ee
We find that the most general $\xi$'s satisfying (\ref{ckv}) in the limit $r\to\infty$ while also maintaining a finite first order effective action $S^{(1)}[\xi]$ (\ref{S1}) are given by
\begin{eqnarray}
\xi^\tau&=&\epsilon(\tau)+\xi^\tau_{-2}(\tau,\theta,\phi)\frac{1}{r^2}+{\cal O}(\frac{1}{r^3})\,,\nn\\
\xi^r&=&-r\epsilon'(\tau)+\xi^r_0(\tau,\theta,\phi)+\xi^r_{-1}(\tau,\theta,\phi)\frac{1}{r}+{\cal O}(\frac{1}{r^2})\,,\label{xi3}\\
\xi^\theta&=&\xi^\theta_{-1}(\tau,\theta,\phi)\frac{1}{r}+\xi^\theta_{-2}(\tau,\theta,\phi)\frac{1}{r^2}+{\cal O}(\frac{1}{r^3})\,,\nn\\
\xi^\phi&=&\xi^\phi_0(\tau)+\xi^\phi_{-1}(\tau,\theta,\phi)\frac{1}{r}+\xi^\phi_{-2}(\tau,\theta,\phi)\frac{1}{r^2}+{\cal O}(\frac{1}{r^3})\,,\nn
\end{eqnarray}
where $\xi^\tau_{-2}, ~\xi^r_0, ~\xi^r_{-1}, ~\xi^\theta_{-1}, ~\xi^\theta_{-2}, ~\xi^\phi_{-1}, ~\xi^\phi_{-2}$ satisfy the following equations:
\begin{eqnarray}
0&=&A(\theta)\,\left(2\,\xi^r_{0}+2\,{\xi^\phi_{0,\,\tau}}\,+\xi^\phi_{-1,\,\phi}-\xi^\theta_{-1,\,\theta}\right)
+B(\theta)\,\xi^\theta_{-1}+C(\theta)\,\xi^\tau_{-2,\,\phi},\label{ckveq1}\\
0&=&D(\theta)\,\xi^\theta_{-1}+E(\theta)\,\left(4\,\xi^r_{0}-\xi^\phi_{-1,\,\phi}-\xi^\theta_{-1,\,\theta}\right)
+F(\theta)\,{\xi^\phi_{0,\,\tau}}\,,\label{ckveq2}\\
0&=&D(\theta)\,\xi^\theta_{-2}+E(\theta)\,\left(5\xi^r_{-1}-\xi^\phi_{-2,\,\phi}-\xi^\theta_{-2,\,\theta}
+3\,\xi^\tau_{-2,\,\tau}\right)+F(\theta)\,\xi^\phi_{-1,\,\tau}+G(\theta)\,\epsilon'(\tau),\quad\quad~\label{ckveq3}\\
0&=&4\,E(\theta)\,\xi^\tau_{-2,\,\theta}+F(\theta)\,\xi^\phi_{-1,\,\theta}.\label{ckveq4}
\end{eqnarray}
where
\bea
&&A(\theta)=32(3+\cos{2\theta})\sin^2{\theta}\,,\nn\\
&&B(\theta)=24(10\sin{2\theta}-\sin{4\theta})\,,\nn\\
&&C(\theta)=-2(3+\cos{\theta})(3+28\cos{2\theta}+\cos{4\theta})\,,\nn\\
&&D(\theta)=-917\cos{\theta}+449\cos{3\theta}-45\cos{5\theta}+\cos{7\theta}\,,\\
&&E(\theta)=2(5\sin{\theta}+\sin{3\theta})(3+28\cos{2\theta}+\cos{4\theta})\,,\nn\\
&&F(\theta)=-256\sin^2{\theta}(5\sin{\theta}+\sin{3\theta})\,,\nn\\
&&G(\theta)=16(5\sin{\theta}+\sin{3\theta})(3+\cos{2\theta})^2\,.\nn
\eea
Since in principle the equations (\ref{ckveq1}-\ref{ckveq4}) may be solved for $\xi^\tau_{-2}, ~\xi^r_0, ~\xi^r_{-1}, ~\xi^\theta_{-1}, ~\xi^\theta_{-2}, ~\xi^\phi_{-1}, ~\xi^\phi_{-2}$ without imposing constraints on $\epsilon(\tau)$ and $\xi^\phi_0(\tau)$, we find that the $\xi$'s in (\ref{xi3}) are of the same order as those in (\ref{xi2}). However, it is too involved to check the Virasoro algebra explicitly in the present case.

Interestingly, we find that the $\xi$'s in (\ref{xi2}) is compatible with equations (\ref{ckveq1}-\ref{ckveq4}) only when $\xi^r_0(\tau,\theta)=\xi^\theta_{-1}(\tau,\theta)=0$ and $\xi^\phi_0(\tau)=C$, leading to stricter asymptotic Killing vectors:
\begin{eqnarray}
\xi^\tau&=&\epsilon(\tau)+\epsilon''(\tau)\frac{1}{2r^2}+{\cal O}(\frac{1}{r^3})\,,\nn\\
\xi^r&=&-r\epsilon'(\tau)+{\cal O}(\frac{1}{r})\,,\label{xi4}\\
\xi^\theta&=&{\cal O}(\frac{1}{r^2})\,,\nn\\
\xi^\phi&=&C-\epsilon''(\tau)\frac{1}{r}+{\cal O}(\frac{1}{r^2})\,.\nn
\end{eqnarray}
The $\xi$'s in (\ref{xi4}) are relaxed slightly compared to (\ref{asyMatsuo}).
Actually for the asymptotic Killing vector (\ref{asyMatsuo}), one can check straightforwardly that it is incompatible with the equations (\ref{ckveq1}-\ref{ckveq4}) unless $\epsilon(\tau)+\epsilon'''(\tau)=0$, which is inconsistent with the existence of an infinite-dimensional Virasoro algebra.

\section{Central Charge}

Employing the quasi-local charge\cite{Brown:1992br} defined by using the surface energy momentum tensor, it turns out that our asymptotic Killing vectors (\ref{xi2}) give rise to Virasoro algebra with the following central extension term
\be
\frac{1}{2\pi i}\oint d\tau \int d\theta\, d\phi\,\, \mathcal{C}_{m,\,n}(\tau,\theta,\phi)\,,\label{cex}
\ee
in which
\bea
&&\mathcal{C}_{m,\,n}(\tau,\theta,\phi)\nn\\
&&= -\, \frac{J\,\epsilon_m(\tau)}{16 \,\pi} (\, \Omega^2 (\,\Omega\,\Gamma' + 4 \,\Gamma\,\Omega') \, \xi^\theta_{-1} + 2 \,\Gamma \,\Omega^3 \,\xi^r_0 - 4 (\,\Omega\,\Gamma' + \Gamma\,\Omega') \, \xi^r_{0,\,\theta} - 4\,\Gamma\,\Omega\, \xi^r_{0,\,\theta\theta})\nn\\
&&~~~-\,\frac{J\,\epsilon_m(\tau)}{64\,\pi \Lambda \,\Gamma\,\Omega} ( \,4\,\Gamma^2\,\Omega^4( \,\epsilon_n'''(\tau) + 2\, \epsilon_n'(\tau)+ 2 \,\xi^r_{-1} + 2\,\xi^\phi_{-2,\,\phi}) + 4\,\Gamma\,\Omega^3 ( \,\Omega\,\Gamma' + 4 \,\Gamma\,\Omega')\, \xi^\theta_{-2} \nn\\
&&~~~-\, 16 \,\Gamma\,\Omega ( \,\Omega\,\Gamma' +  \,\Gamma\,\Omega') \,\xi^r_{-1,\,\theta} + 16\, \Gamma^2 (\,\Omega^2 - 1) \,\xi^r_{-1,\,\phi\phi} - 16 \,\Gamma^2\,\Omega^2 \,\xi^r_{-1,\,\theta\theta} \nn\\
&&~~~ +\, 4\,\Gamma^2\,\Omega^4 ( \,3\,{\xi^r_0}^2 - {\xi^r_{0,\,\theta}}^2 - 4\,{\xi^\phi_{0,\,\tau}}^2 ) + (\, 4\,\Gamma^2\,\Omega^4 + 16 \,\Gamma^2\,\Omega'^{\,2} + 16 \,\Gamma^2\,\Omega\,\Omega'' + 16 \,\Gamma\,\Omega\,\Gamma'\,\Omega' \nn\\
&&~~~-\, 8 \,\Gamma\,\Omega^3\,\Gamma'\,\Omega' - \Omega^4\,\Gamma'^{\,2}) \,{ \xi^\theta_{-1}}^2 + 16\,\Gamma\,\Omega (\,\Omega^3\,\Gamma' - 3\,\Omega\,\Gamma' - 3\,\Gamma\,\Omega') \,\xi^r_0\,\xi^r_{0,\,\theta}\nn\\
&&~~~ +\, 16 \,\Gamma^2\,\Omega^2 (\,\Omega^2 - 3)  \,\xi^r_0\,\xi^r_{0,\,\theta\theta} - 16 \,\Gamma\,\Omega (\,\Omega\,\Gamma' + \,\Gamma\,\Omega')\,\xi^\theta_{-1}\, \xi^\theta_{-1,\,\theta} - 16 \,\Gamma^2\,\Omega^2 \,\xi^\theta_{-1}\, \xi^\theta_{-1,\,\theta\theta}\nn\\
&&~~~ +\, 4 \,\Gamma\,\Omega^3 (\,\Omega\,\Gamma' + 8\,\Gamma\,\Omega')\, \xi^\theta_{-1}\,\xi^r_0 + 8 (\,\Omega^2\,\Gamma'^{\,2} + 2\,\Gamma^2\,\Omega'^{\,2} + \,\Gamma\,\Omega\,\Gamma'\,\Omega' - 2\,\Gamma^2\,\Omega\,\Omega'')\, \xi^\theta_{-1}\, \xi^r_{0,\,\theta}\nn\\
&&~~~ +\, 8\,\Gamma\,\Omega^2\,\Gamma'\,\xi^\theta_{-1}\, \xi^r_{0,\,\theta\theta} + 16 \,\Gamma\,\Omega (\,2\,\Omega\,\Gamma' + \Gamma\,\Omega')\, \xi^\theta_{-1,\,\theta}\, \xi^r_{0,\,\theta} + 32\,\Gamma^2\,\Omega^2\,\xi^\theta_{-1,\,\theta}\, \xi^r_{0,\,\theta\theta} \nn\\
&&~~~ +\, 16 \,\Gamma^2\,\Omega^2\,\xi^\theta_{-1,\,\theta\theta}\, \xi^r_{0,\,\theta} + 8 \,\Gamma\,\Omega^3(\,(\,\Omega\,\Gamma' + 2\,\Gamma\,\Omega')\,\xi^\theta_{-1} + 2\,\Omega\,\Gamma'\, \xi^r_{0,\,\theta} - 2\,\Gamma\,\Omega \, \xi^r_0 + 2\,\Gamma\,\Omega \,\xi^r_{0,\,\theta\theta})\,\xi^\phi_{0,\,\tau})\,,\nn\\
&&
\eea
where $\epsilon_m(\tau) = \tau^{1+m}\,,~~\epsilon_n(\tau) = \tau^{1+n}\,$, \,$\Lambda$ is the regularization constant and
\be
\Gamma \equiv {1+\cos^2\theta\over 2}\;, \;\;\;\;\; \Omega \equiv {2 \sin \theta\over 1+\cos^2\theta}\,.
\ee
Integration (\ref{cex}) is defined by considering the analytic continuation of $\tau$. The central charge can be read off from the central extension term (\ref{cex}), which should be of the following form in principle:
\be
\frac{c}{12}\,m(m^2 - \alpha) \delta_{m+n,\,0}\,.
\ee

\vspace*{5mm}



\begin{thebibliography}{99}


\bibitem{Guica:2008mu}
  M.~Guica, T.~Hartman, W.~Song and A.~Strominger,
  ``The Kerr/CFT Correspondence,''
  Phys.\ Rev.\  D {\bf 80}, 124008 (2009)
  [arXiv:0809.4266 [hep-th]].

\bibitem{Brown:1986nw}
  J.~D.~Brown and M.~Henneaux,
  ``Central Charges in the Canonical Realization of Asymptotic Symmetries: An
  Example from Three-Dimensional Gravity,''
  Commun.\ Math.\ Phys.\  {\bf 104}, 207 (1986).

\bibitem{Matsuo:2009sj}
  Y.~Matsuo, T.~Tsukioka and C.~M.~Yoo,
  ``Another Realization of Kerr/CFT Correspondence,''
  Nucl.\ Phys.\  B {\bf 825}, 231 (2010)
  [arXiv:0907.0303 [hep-th]].

\bibitem{Castro:2009jf}
  A.~Castro and F.~Larsen,
  ``Near Extremal Kerr Entropy from AdS$_2$ Quantum Gravity,''
  JHEP {\bf 0912}, 037 (2009)
  [arXiv:0908.1121 [hep-th]].

\bibitem{Matsuo:2009yet}
  Y.~Matsuo, T.~Tsukioka and C.~M.~Yoo,
  ``Yet Another Realization of Kerr/CFT Correspondence,''
  Europhys.\ Lett.\  {\bf 89} (2010) 60001
  [arXiv:0907.4272 [hep-th]].

\bibitem{Matsuo:2010newlimit}
  Y.~Matsuo and T.~Nishioka,
  ``New Near Horizon Limit in Kerr/CFT,''
  JHEP {\bf 1012}, 073 (2010)
  [arXiv:1010.4549 [hep-th]].

\bibitem{Porfyriadis:2010vg}
  A.~P.~Porfyriadis and F.~Wilczek,
  ``Effective Action, Boundary Conditions, and Virasoro Algebra for AdS$_3$,''
  [arXiv:1007.1031 [gr-qc]].

\bibitem{Barnich:2001jy}
  G.~Barnich and F.~Brandt,
  ``Covariant theory of asymptotic symmetries, conservation laws and  central
  charges,''
  Nucl.\ Phys.\  B {\bf 633}, 3 (2002)
  [arXiv:hep-th/0111246].


\bibitem{Barnich:2007bf}
  G.~Barnich and G.~Compere,
  ``Surface charge algebra in gauge theories and thermodynamic integrability,''
  J.\ Math.\ Phys.\  {\bf 49}, 042901 (2008)
  [arXiv:0708.2378 [gr-qc]].

\bibitem{Brown:1992br}
  J.~D.~Brown and J.~W.~.~York,
  ``Quasilocal energy and conserved charges derived from the gravitational
  action,''
  Phys.\ Rev.\  D {\bf 47}, 1407 (1993)
  [arXiv:gr-qc/9209012].


\end{thebibliography}
\end{document}